\documentclass[referee]{jfm}

\usepackage{graphicx}
\usepackage{natbib}
\usepackage{subfig}
\usepackage{psfrag}
\usepackage{color}
\usepackage{epstopdf}

\ifCUPmtlplainloaded \else
  \checkfont{eurm10}
  \iffontfound
    \IfFileExists{upmath.sty}
      {\typeout{^^JFound AMS Euler Roman fonts on the system,
                   using the 'upmath' package.^^J}%
       \usepackage{upmath}}
      {\typeout{^^JFound AMS Euler Roman fonts on the system, but you
                   dont seem to have the}%
       \typeout{'upmath' package installed. JFM.cls can take advantage
                 of these fonts,^^Jif you use 'upmath' package.^^J}%
      }
  \else
  \fi
\fi

\ifCUPmtlplainloaded \else
  \checkfont{msam10}
  \iffontfound
    \IfFileExists{amssymb.sty}
      {\typeout{^^JFound AMS Symbol fonts on the system, using the
                'amssymb' package.^^J}%
       \usepackage{amssymb}%
         \let\leq=\leqslant
         \let\geq=\geqslant
      }{}
  \fi
\fi

\ifCUPmtlplainloaded \else
  \IfFileExists{amsbsy.sty}
    {\typeout{^^JFound the 'amsbsy' package on the system, using it.^^J}%
     \usepackage{amsbsy}}
    {\providecommand\boldsymbol[1]{\mbox{\boldmath $##1$}}}
\fi

\newsavebox{\astrutbox}
\sbox{\astrutbox}{\rule[-5pt]{0pt}{20pt}}

\title[On bubble clustering  and energy spectra in pseudo-turbulence]{On bubble clustering and energy spectra in pseudo-turbulence}

\author[J. Mart\'inez, D. Chehata, D.P.M. van Gils, C. Sun, and D. Lohse]%
{J\ls U\ls L\ls I\ls \'A\ls N\ns M\ls A\ls R\ls T\ls \'I\ls N\ls E\ls Z\ns M\ls E\ls R\ls C\ls A\ls D\ls O,\ns \break
D\ls A\ls N\ls I\ls E\ls L\ns  C\ls H\ls E\ls H\ls A\ls T\ls A\ns G\ls \'O\ls M\ls E\ls Z,\ns \break
D\ls E\ls N\ls N\ls I\ls S\ns V\ls A\ls N\ns G\ls I\ls L\ls S,\ns  \break
C\ls H\ls A\ls O\ns S\ls U\ls N,\ns \and D\ls E\ls T\ls L\ls E\ls F\ns L\ls O\ls H\ls S\ls E\ns}

\affiliation{Physics of Fluids Group, Department of Science and Technology, J.M. Burgers Center for Fluid Dynamics, and  IMPACT Institute, University of Twente, PO Box 217, 7500 AE  Enschede, The Netherlands}

\pubyear{1996}
\volume{538}
\pagerange{119--126}
\date{30th September  2009}

\begin{document}

\maketitle

\begin{abstract}

3D-Particle Tracking (3D-PTV) and Phase Sensitive Constant Temperature Anemometry in pseudo-turbulence---i.e., flow solely driven by rising bubbles--- were performed to investigate bubble clustering and to obtain the mean bubble rise velocity, distributions of bubble velocities, and energy spectra at  dilute gas concentrations ($\alpha \leq2.2$\%). To characterize the clustering the pair correlation function $G(r,\theta)$ was calculated. The  deformable bubbles with equivalent bubble diameter $d_b=4-5$ mm were found to cluster within a radial distance of a few bubble radii with a preferred vertical orientation.  This vertical alignment was present at both small and large scales.  For small distances also some horizontal clustering was found. The large number of data-points and the non intrusiveness of PTV  allowed to obtain  well-converged Probability Density Functions (PDFs) of the bubble velocity. The PDFs had a non-Gaussian form for all velocity components and intermittency effects could  be observed. The energy spectrum of the liquid velocity fluctuations  decayed with a power law of $-3.2$, different from the $\approx -5/3$ found for homogeneous isotropic turbulence, but close to the prediction $-3$ by \cite{lance} for pseudo-turbulence.

\end{abstract}

\section{Introduction}\label{sec:intro}

Bubbly pseudo-turbulence---i.e. a flow solely driven by rising bubbles---is relevant from an application point of view due to the omnipresence of bubble columns, e.g. in the chemical industry, in water treatment plants, and in the steel industry (\cite{bubble:columns}). A better understanding of the involved phenomena is necessary for scaling-up industrial devices and for optimization and performance prediction. This article wants to provide experimental data on pseudo-turbulence by means of novel experimental techniques. The main questions to be addressed are:

\begin{itemize}
\item[(i)] What is the preferential range and the orientation of bubble clustering in pseudo-turbulence?
\item[(ii)] What is the mean bubble rise velocity and what kind of probability distribution does the bubble velocity have?
\item[(iii)] What is the shape of the energy spectrum of pseudo-turbulence?
\end{itemize}

\subsection{Bubble clustering}
In dispersed flows, the hydrodynamic interaction between the two-phases  and  the particle inertia result in an inhomogeneous distribution of both particles and bubbles (see e.g. experimentally \cite[][]{ayyala,salazar,saw} and numerically \cite[][]{enrico:2,enrico,cencini} and
\cite{toschi} for a general recent review). Bubble clustering in \emph{pseudo-turbulence} has been studied numerically \cite[][]{smereka,sangani,kazu,mazzitelli:2} and experimentally \cite[][]{cartellier,risso,zenit,roig}. The numerical work by \cite{smereka} and by \cite{sangani} and theoretical work by \cite{wijngaarden}, \cite{kok}, and \cite{wijngaarden2}  suggest that, when  assuming  potential flow, rising bubbles form  horizontally aligned rafts. Three dimensional direct numerical simulations, which also solve the motion of the gas-liquid interface at the bubble's surface, have  become available in the last few years. The work by \cite{Trygg:Ia,Trygg:II} suggests that deformability effect plays  a crucial role for determining the orientation of the clustering. For spherical non-deformable bubbles these authors simulated  up to 216 bubbles with  Reynolds numbers in the range of $12-30$ and void fraction $\alpha$ up to $24\%$, and Weber number of about 1. For the deformable case, they simulated  27 bubbles with Reynolds number of 26, Weber number of 3.7, and $\alpha=6\%$. The authors found that the orientation of bubble clusters is strongly influenced by the deformability of the bubbles: spherical pairs of bubbles have a high probability to align horizontally, forming rafts, whereas the non-spherical ones  preferentially align in the vertical orientation. In a later investigation, where  inertial effects were more pronounced, \cite{Trygg:III} studied both cases  for bubble Reynolds number of  order 100. In this case only a weak vertical cluster was  observed in a swarm of  14 deformable bubbles. Their explanation for the weaker vertical clustering was that the wobbly bubble motion, enhanced by the larger Reynolds number, produces perturbations which do not allow the bubbles to align vertically and accelerate the break up  in case some of them cluster.

In spite of numerous experimental studies on pseudo-turbulence, bubble clustering has not yet been fully quantitatively analysed experimentally. \cite{zenit} found a mild horizontal clustering using 2D imaging techniques for bubbles with particulate Reynolds number higher than 100. \cite{cartellier} studied the microstructure  of homogeneous bubbly flows for Reynolds number of order 10. They found a moderate horizontal accumulation using pair density measurements with two optical probes. Their results showed a higher probability of the pair density in the horizontal plane and a reduced bubble density behind the wake of a test bubble. \cite{risso} performed experiments with a swarm of deformable bubbles ($d_b$=2.5 mm), aspect ratio around 2, and Reynolds number of 800. They did not find  clustering and suggested that  in this low void fraction regime  ($\alpha<1.05\%$) there was a weak influence of hydrodynamic interaction between bubbles.

\subsection{Mean bubble rise velocity and statistics of bubble velocity}
The mean bubble rise velocity in bubbly flows is found to decrease with increasing bubble concentration whereas the normalized vertical fluctuation $V_{z,rms}/\overline{V}_z$ increases \cite[][]{zenit,Trygg:II,julian}. The interpretation is that when increasing the concentration the bubble-induced counterflow becomes more important and in addition the
hydrodynamic interactions between bubbles  become more frequent and hinder the upward movement of bubbles, provoking at the same time,  an increment of the bubble velocity fluctuations.

 Next, Probability Density Functions (PDFs) of bubble velocities provide useful information for effective force correlations used in bubbly flow models in industry. PDFs in pseudo-turbulence have been obtained in  the numerical studies of \cite{Trygg:Ib,Trygg:II} and \cite{Trygg:III}.  For non-deformable bubbles \cite[]{Trygg:Ib}, the PDFs of velocity fluctuations have a Gaussian distribution. If deformability is considered \cite[]{Trygg:II}, the PDF of only one horizontal component of the velocity keeps a Gaussian shape while the non-Gaussianity in the PDFs was stronger at the lowest concentration $\alpha=2\%$, recovering a Gaussian distribution for $\alpha=6\%$. However in that numerical work only a limited number of bubbles ($N_b=27$) could be considered and the different behavior in the two different horizontal directions reflect the lack of statistical convergence. Experimental PDFs of bubble velocities in pseudo-turbulence have been obtained  by \cite{zenit} and by \cite{julian}. In these studies the bubble velocity was measured intrusively using an impedance probe technique. The  amount of data-points used for the PDFs was not sufficient for good statistical convergence.

\subsection{Liquid energy spectrum}

In bubbly flows, one must distinguish between the energy input due to the bubbles and the energy input due to some external forcing---either of them can be predominant. \cite{lance} suggested  the ratio of the bubble-induced kinetic energy and the kinetic energy of the flow without bubbles as appropriate dimensionless number to characterize the type of the bubbly flow.  \cite{judith} called this ratio the  \emph{bubblance} parameter $b$, defined as

\begin{equation}
b=\frac{1}{2} \frac{\alpha U^2_r}{u^{\prime2}_0}, \label{eq:bubblance}
\end{equation}

\noindent where $\alpha$ was already defined, $U_r$ is the rise  bubble velocity in still water, and $u^{\prime}_0$ is the typical turbulent vertical fluctuation in the absence of bubbles. \cite{lance} measured the liquid power spectrum in bubbly turbulence and observed a gradual change of the slope with increasing void fraction. At low concentrations the slope of the spectrum was close to the Kolmogorov's  turbulent value of $-5/3$. By increasing $\alpha$, the  principal driving mechanism changed---the forcing now more and more originated from the bubbles and not from some external driving. In that regime the slope was close to $-8/3$. Having in mind equation~(\ref{eq:bubblance}) one may expect different scaling behavior, depending on the nature of the energy input that is dominant, namely  $b<1$ for dominant turbulent fluctuations or $b>1$ for dominant bubble-induced fluctuations. Indeed from table 1 of \cite{judith} one may get the conclusion that the slope is around $-5/3$ for $b<1$ and around $-8/3$ for $b>1$. Also \cite{riboux} obtained a spectral lope of about $-$3 in the wake of a swarm of  rising bubbles in still water ($b=\infty$). Moreover, in numerical simulations \cite{kazu} obtained the same spectral slope for the velocity fluctuations caused by up to 800 rising light particles, i.e. $b=\infty$, with finite diameter ($Re\approx30$). However, there are also counter-examples: e.g., \cite{mudde}, and \cite{cui} obtained around $-5/3$, though $b=\infty$. Therefore, in this paper we want to re-examine the issue of the spectral slope in pseudo-turbulence ($b=\infty$).

\subsection{Outline of paper}
The paper is structured as follows: in section \ref{exper} the experimental apparatus is explained and Particle Tracking Velocimetry technique, and Phase Sensitive Constant Temperature Anemometry are described. Section \ref{res} is divided in three sub-sections: in the first part results on bubble clustering are shown, followed by the results on the mean bubble rise  velocity and the bubble velocity distributions, and finally the power spectrum measurements are presented. Finally, a summary and an outlook on future work  are presented in section \ref{conclusions}.

\section{Setup, tools, and methods} \label{exper}

\subsection{Twente water channel}
\begin{figure}
\centering
\includegraphics[height=8cm]{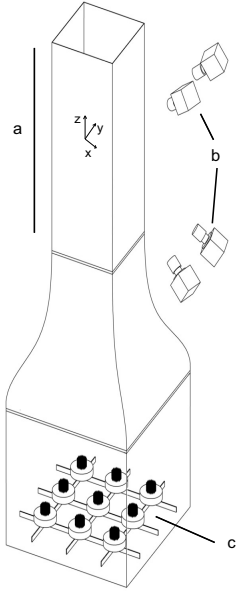}
\caption{Experimental apparatus: a) Measurement section of length 2m, b) 4-camera PTV system, c) Capillary islands.} \label{fig:diagram_tunnel}
\end{figure}

\begin{figure}
\centering
\includegraphics[width=8cm]{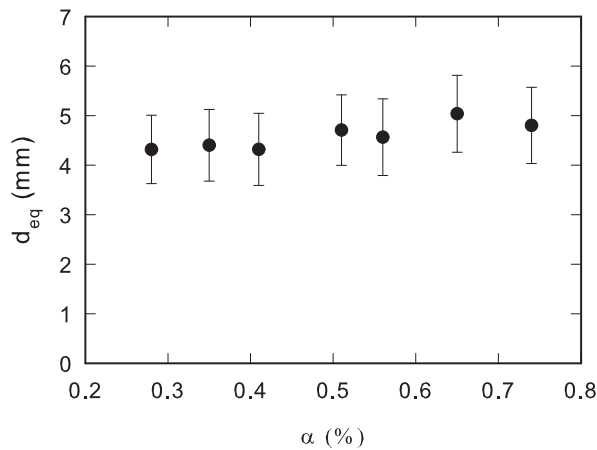}
\caption{Equivalent bubble diameter $d_{eq}$ as a function of void fraction $\alpha$. Standard deviation as error bars. We measured the long $d_l$ and short axis $d_s$ of 400 bubbles per concentration from 2D images. The equivalent diameter was obtained by assuming ellipsoidal bubbles with a  volume equal to that of a  spherical bubble, $d_{eq}=(d_s d_l^2)^{1/3}$
}
\label{fig:bubdiam}
\end{figure}

The experimental apparatus consists of  a vertical water tunnel with a  2 meter long measurement section with 0.45 $\times$ 0.45 m$^2$ cross section. A sketch depicting the setup is shown in figure \ref{fig:diagram_tunnel}. The measurement section has three glass walls for optical access and illumination (see \cite{judith} for more details).  The  channel was filled with deionized water until the top of the measurement section. The level of the liquid column was 3.8 m above the place where bubbles were injected. We used 3 capillary islands  in the lowest part of the channel to generate bubbles. Each island contains 69 capillaries with an inner diameter of 500 $\mu$m. Different bubble concentrations were achieved varying the air flow through the capillary islands. We performed experiments with dilute bubbly flows with typical void fractions in the range $0.28\% \leq \alpha \leq 0.74\%$ for PTV and  in the range $0.20\% \leq \alpha \leq 2.2\%$ for CTA. The void fraction $\alpha$ was determined using an U-tube manometer which measures the pressure difference between two points at different heights of the measurement section  \cite[see][]{judith}.
A mono-dispersed  bubbly swarm with mean bubble diameter of $4-5$ mm was studied. Typical Reynolds numbers $Re$ are of the order $10^3$, the Weber number $We$ is in the range 2$-$3  (implying deformable bubbles) and E\"otvos number $Eo$ around $3-4$. Table~\ref{t:exp_par} defines these various dimensionless numbers and summarizes the experimental conditions. In figure~\ref{fig:bubdiam}, the mean bubble diameter $d_{eq}$ as a function of void fraction is shown. The values are within the range of 4-5 mm and show a slight increment with bubble concentration.

\begin{table}
\begin{center}

\begin{tabular}{cccccc}

$\alpha (\%)$&~$d_{b}$(mm)&~~$\overline{V}_{z}$(m/s)&~Re&~We&~Eo
\\ \hline
$0.28-0.74$&~$4-5$&~~0.16$-$0.22&~1000&~2$-$3&~3$-$4\\

\end{tabular}
\caption{Typical experimental parameters: void fraction $\alpha$, mean bubble diameter ($d_b=\sqrt[3]{d_l^2d_s}$), $d_l$ and $d_s$ are the long and short axis of 2D image of a given ellipsoidal bubble, mean rising velocity in still water $\overline{V}_{z}$, and Reynolds ($Re=d_b \overline{V}_z /\nu_f$), Weber ($We=\rho d_b \overline{V}^2_z/\sigma$), and E\"otvos ($Eo=\rho_f g d_b^2/ \sigma$) numbers. Here $\nu_f=1\times10^{-6}\ m^2/s$ is the kinematic viscosity of water, $\sigma=0.072\ N/m$ the surface tension at the air-water interface, and $g$ the gravitational acceleration.} \label{t:exp_par}
\end{center}
\end{table}

\subsection{Particle Tracking Velocimetry} \label{sec:ptv}
In the last few years 3D-Particle Tracking Velocimetry (PTV) has become a powerful measurement technique in fluid mechanics. The rapid development of high-speed imaging has enabled a  successful implementation of the technique in studies on turbulent motion of particles \cite[e.g.][]{pinton1,luethi,bodenschatz,berg,pinton2,toschi}. The measured 3D spatial position of particles and time trajectories allow for a Lagrangian description which  is the natural approach  for transport mechanisms.

Figure~\ref{fig:diagram_tunnel} also sketches the positions of the four high-speed cameras (Photron 1024-PCI) which were used to image the bubbly flow. The four cameras were viewing from one side of the water channel and were focused in its central region, at a height of 2.8 m above the capillary islands. Lenses with 50 mm focal length were attached to the cameras. We had a  depth of field  of 6 cm. The image sampling frequency was 1000 Hz using a camera resolution of 1024$\times$1024 pixel$^2$. The cameras were triggered externally in order to achieve a fully synchronized acquisition.  We used the PTV software developed at IfU-ETH for camera calibration and tracking algorithms. For a detailed description of this technique we refer to the work of \cite{hoyer} and references therein. A 3D solid target was used for calibration. Bubbles were detected within a volume of 16$\times$16$\times$6 cm$^3$ with an accuracy of 400 $\mu$m.  To illuminate the measuring volume homogeneous back-light and a diffusive plate were used in order to get the bubble's contour imaged as a dark shadow. The image sequence  was binarized after subtracting a sequence-averaged background; then these images were used along with the PTV software to get the 3D positions of the bubbles. We acquired 6400 images per camera corresponding to 6.4 s of measurement (6.7 Gbyte image files).

For higher bubble concentration, many bubbles are imaged as merged blobs and can not be identified as individual objects. These merged bubbles images are not considered for the analysis. Therefore the number $N_b$ of clearly identified individual bubbles goes down with increasing void fraction. For the most dilute case ($\alpha=0.28\%$) around $N_b\approx190$ bubbles were detected in each image. This quantity dropped to nearly 100 for the most concentrated cases ($\alpha=0.65\%$ and $\alpha=0.74\%$ ). If a bubble is tracked in at least 3 consecutive time-steps, we call it a linked bubble. Table~\ref{t:imag_stat} summarizes typical values of number of bubbles ($N_b$) and linked bubbles ($N_{link}$) per image.

\begin{table}
\begin{center}

\begin{tabular}{cccc}

$\alpha$&~~$N_b$&~~$N_{link}$&~~$N_{link}/N_b\cdot100\%$
\\ \hline
0.28\%&~~$\approx$190&~~$\approx$50&~~27\%\\
0.35\%&~~$\approx$190&~~$\approx$50&~~27\%\\
0.41\%&~~$\approx$170&~~$\approx$40&~~24\%\\
0.51\%&~~$\approx$140&~~$\approx$30&~~21\%\\
0.56\%&~~$\approx$140&~~$\approx$30&~~21\%\\
0.65\%&~~$\approx$110&~~$\approx$20&~~18\%\\
0.74\%&~~$\approx$110&~~$\approx$15&~~14\%\\

\end{tabular}
\caption{Number of detected bubbles and linked bubbles per image for all the concentration studied. For the highest void fractions there is a drop in the number of linked bubbles, since most bubbles are imaged as 2D merged blobs, which are not considered in the analysis.}
\label{t:imag_stat}
\end{center}
\end{table}

\subsection{Pair correlation function}

\begin{figure}
\centering
\includegraphics[width=0.45\textwidth]{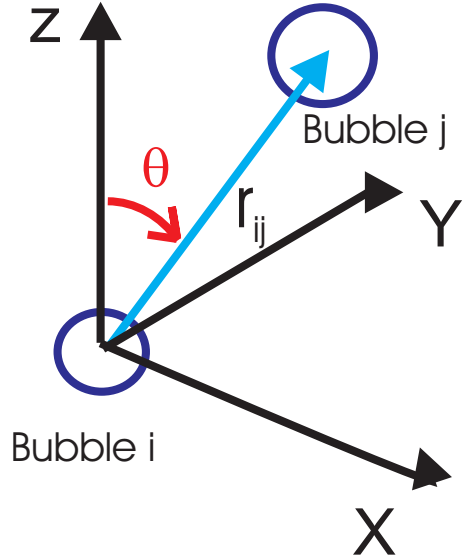}
\caption{Definition of angle $\theta$ between a pair of neighboring bubbles.} \label{fig:def_angle}
\end{figure}

Particle clustering can be quantified using different mathematical tools like pair correlation functions \cite[][]{Trygg:Ia}, Lyapunov exponents \cite[][]{bec2}, Minkowski functionals \cite[][]{enrico},  or PDFs of the distance of two consecutive bubbles in a time-series \cite[][]{enrico:2}. In this investigation the pair correlation function $G(r,\theta)$ is employed to understand how the bubbles are globally distributed. It is defined as follows:

\begin{equation}
G(r,\theta)=\frac{V}{N_b(N_b-1)}\Bigg \langle \sum_{i=1}^{N_b}\sum_{j=1,i\neq j}^{N_b}\delta(\mathbf{r}-\mathbf{r}_{ij}) \Bigg \rangle,
\label{eq:G}
\end{equation}

\noindent where $V$ is the size of the calibrated volume, $N_b$ is the number of bubbles within that volume, $\mathbf{r}_{ij}$ is the vector linking the centers of bubble i and bubble j, and $\mathbf{r}$ is a vector with magnitude $r$ and orientation $\theta$, defined as the angle between the vertical unit vector and the  vector linking  the centers of bubbles $i$ and $j$, as shown in figure~\ref{fig:def_angle}.  From (\ref{eq:G}), the radial and angular pair probability functions can be derived. To obtain the radial pair probability distribution function $G(r)$ one must integrate over spherical shells of radius $r$ and width $\Delta r$, whereas for the angular pair probability distribution function $G(\theta)$ an r-integration is performed.

\subsection{Phase sensitive Constant Temperature Anemometry}\label{sec:cta}

Hot-film measurements in bubbly flows impose considerable difficulty due to the fact that liquid and gas information is present in the signal. The challenge is to distinguish and classify  the signal corresponding to each phase. The hot-film probe does not provide by itself  means for a successful identification. Thus many parametric and non-parametric signal processing algorithms have been used to separate the information of both phases, i.e. thresholding \cite[][]{bruun} or pattern recognition methods  \cite[][]{luther}. The output of these algorithms is a constructed indicator function which labels the liquid and gas parts of the signal. We follow an alternative way and \emph{measure} the indicator function, therefore avoiding uncertainties when computing it. This can be achieved by combining optical fibers for phase detection to the hot-film sensor \cite[see][]{cartellier2,mudde2}.

\begin{figure}
\psfrag{A}{hot-film}
\psfrag{B}{optical fiber}
\centering
\includegraphics[width=0.5\textwidth]{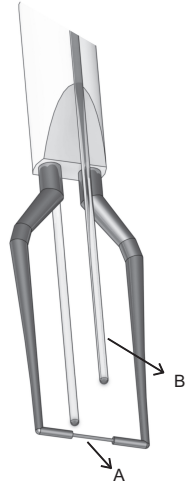}
\caption{Phase Sensitive CTA probe with two attached optical fibers for bubble-probe detection.} \label{fig:CTA_probe}
\end{figure}

The device, called Phase Sensitive CTA, was firstly developed by \cite{ramon:2}. In this technique,  an optical fiber is attached close to the hot-film probe so that when a bubble impinges onto the sensor it also interacts with the optical fiber. Its working principle is based on the different index of refraction of gas and liquid. A light source is coupled into the fiber. A photodiode measures the intensity of the light that is reflected from the fiber tip via a fiber coupler. The incident light leaves the tip of the fiber when immersed in water, but it is reflected back into the fiber  when exposed to air, implying a sudden rise in the optical signal. Thus the intensity of the reflected light indicates the presence of either  gas or liquid at the fiber's tip. In this way, if both signals are acquired simultaneously, the bubble collisions can straightforwardly be detected  without the need of any signal-processing method. For the construction of the probe we used a  DANTEC cylindrical probe (55R11)  and attached  two optical fibers to it (Thorlabs NA=0.22 and 200 $\mu$m  diameter core). We used two fibers in order to assure the detection of all bubbles interacting with the probe. The fibers were glued and positioned at the side of the cylindrical hot-film, at a distance of about 1 mm from the hot-film. An illustration of the hot-film and fibers is shown in figure~\ref{fig:CTA_probe}. The probe was put in the center of the measurement section positioning its supporting arm  parallel to the vertical rising direction of the bubbles so that the axis of the optical probes are also aligned with the preferential direction of the flow, thus allowing for a better bubble-probe interaction and aiming at minimize the  slow down of bubbles approaching the probe. \cite{ramon:2} measured flow velocity with and without the fiber being attached to the probe. They found that the presence of the fibers do not compromise the probe's bandwidth and that its influence on the power spectrum is negligible.

With the signal of each optical fiber a discretized phase indicator function $\xi=\{\xi_i\}_{i=1}^{N}$ can be constructed, whose definition follows  \cite{bruun}, namely

\begin{equation}
\xi_i=\left\{
\begin{array}{ll}
1 & \mathrm{liquid}, \\
0 & \mathrm{bubble}.
\end{array}
 \right.
 \label{eq:if}
\end{equation}

\begin{figure}
  \centering
  \subfloat{\label{fig:typsig}\includegraphics[width=0.5\textwidth]{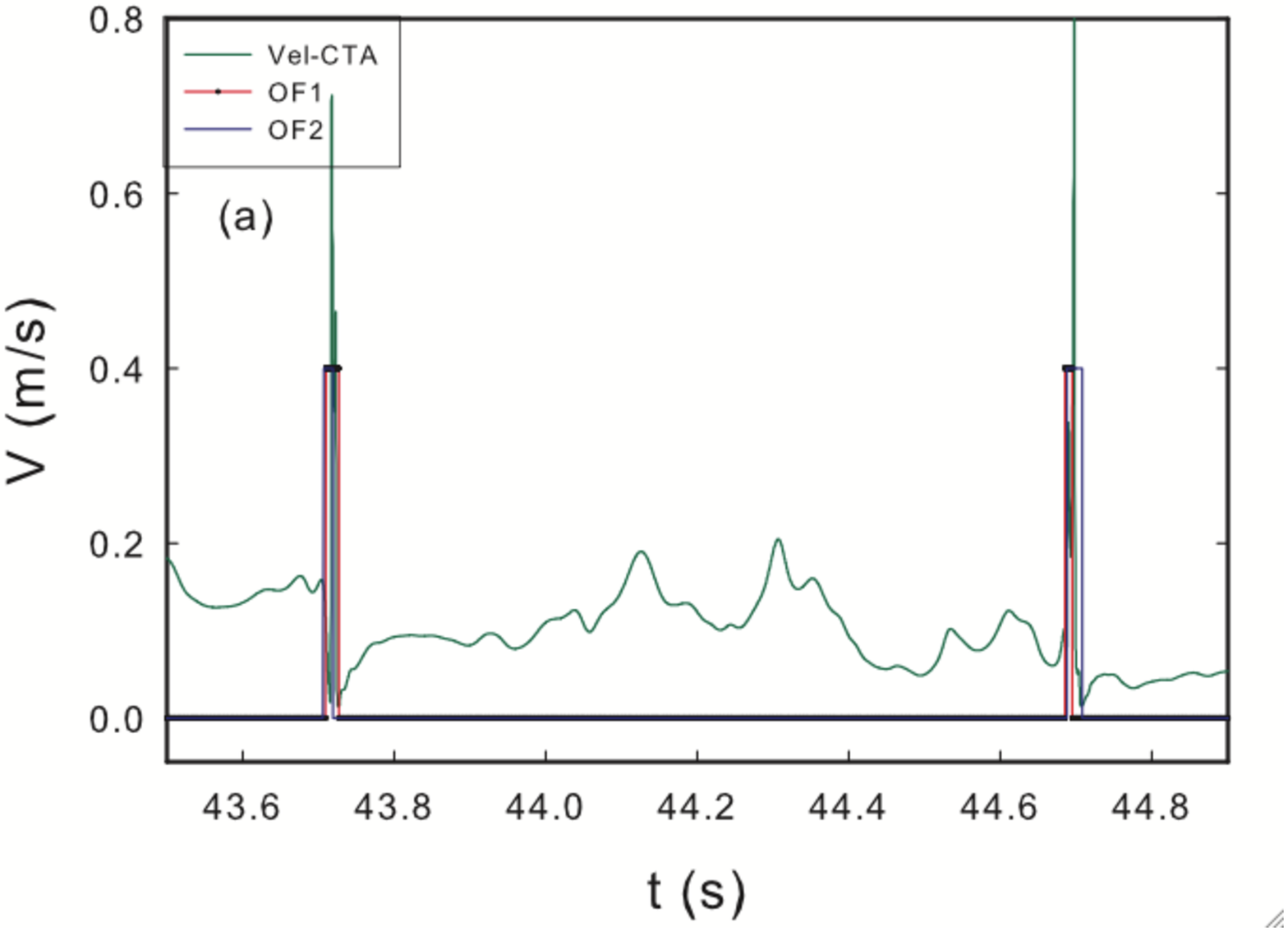}}
  \subfloat{\label{fig:typsig_zoom}\includegraphics[width=0.5\textwidth]{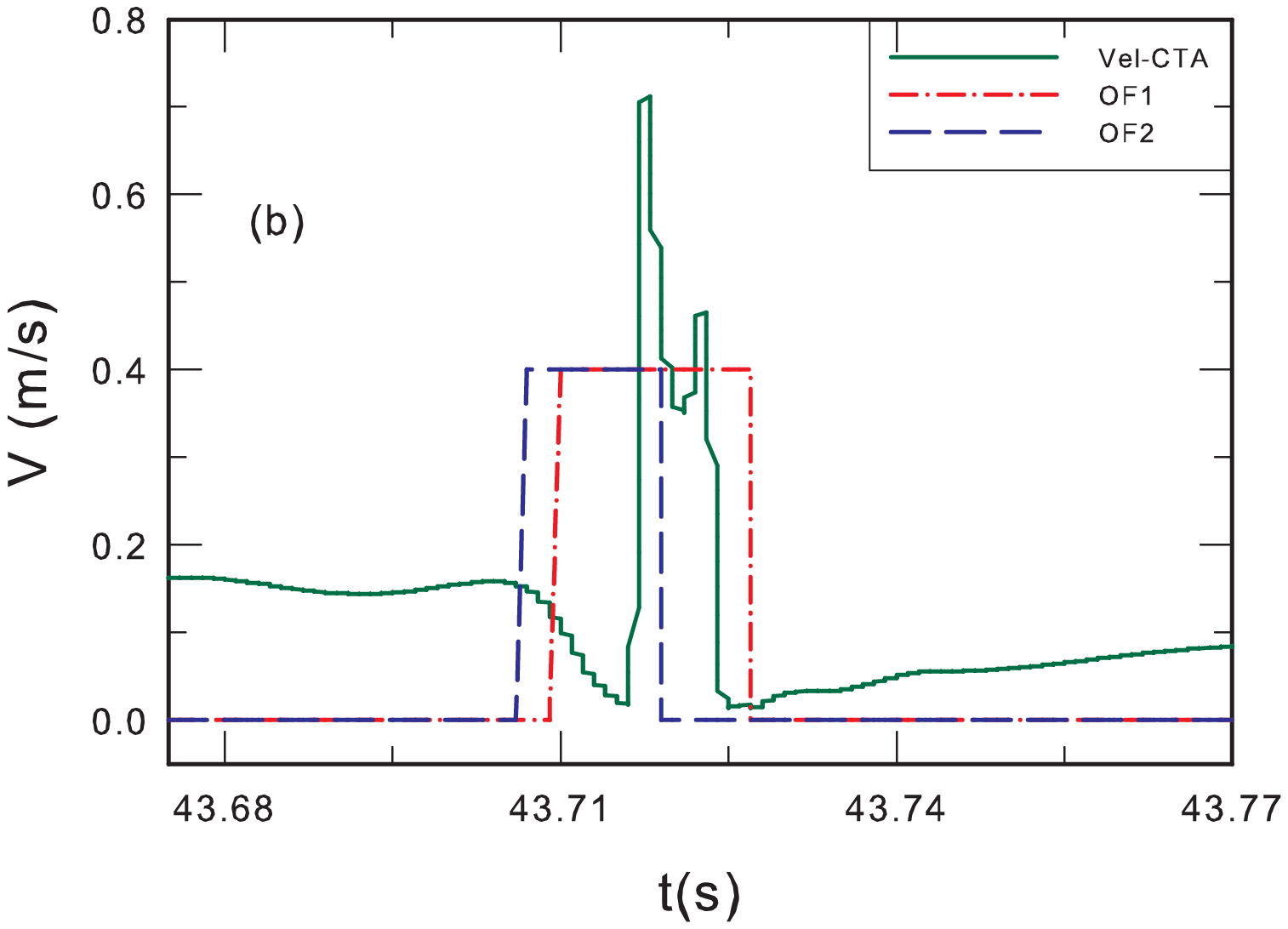}}
  \caption{Typical phase Sensitive CTA signal already transformed into velocity. (a) two consecutive bubble-probe interactions for $\alpha$=0.2\%, (b) zoom in of a bubble collision.}
  \label{fig:CTA}
\end{figure}

An unified phase indicator function is then obtained by multiplying the  indicator functions of both fibers. A typical signal of Phase Sensitive CTA  with two consecutive bubble-probe interactions is shown in figure~\ref{fig:typsig}.
There is one adjustable parameter to construct the indicator functions of the fibers. As explained above the phase discrimination can  be done by an intensity threshold of the optical fiber's signal. When the rising edge of the signal surpasses this threshold,  the indicator function of the fiber must change from a value 1 to 0, as defined in equation~\ref{eq:if}. If the optical fiber was positioned at the same place of the hot-film probe, the rising edge of its signal would occur precisely at the time when the bubble interacts with the hot-film. However, there is a separation of about 1 mm between the fibers and the hot-film, so that the rising edge of the optical fibers' signal occurs actually with some delay. Therefore, to construct the phase indicator function of the optical fibers, one must account for this delay and define the beginning of the interaction not where the signal surpasses the intensity threshold but some time before. The time used for this shifting was obtained considering the vertical separation  between the optical fibers and the hot-film probe and mean bubble velocities: in our experiment a bubble travels 1 mm in about 5 to 7 ms. We shifted the beginning of the collision 8 ms prior to the optical fibers' signal starts to rise from its base value. The shifting value was double checked by analyzing the histograms of the duration of bubble collisions. As it can be observed from figure~\ref{fig:typsig_zoom}, with this shift sometimes part of the CTA signal when the bubbles is approaching the probe ($<10$ ms) was lost, but we noticed no effect on the spectrum when varying the shift duration, provided the bubble spikes were still removed.

With the phase indicator function only pieces containing liquid information of the time series are used for further analysis--- i.e.the part of the signal between the two bubble interactions shown in figure~\ref{fig:typsig}. For each part of the time-series containing liquid information the spectrum was calculated. Then all spectra were averaged and the power spectrum for each bubble concentration was obtained. In this way neither an interpolation nor auto-regressive models for gapped time series were necessary. In our experiments the gas volume fractions varied from $0.2\leq\alpha\leq2.2\%$. The CTA was calibrated using a DANTEC LDA system (57N20 BSA). The calibration curve was obtained by fitting the data points to a 4th order polynomial. The total measuring time was 1 hour for each concentration and the sampling rate of the hot-film and optical fibers was 10 kHz.

\section{Bubble clustering and distributions in pseudo-turbulence} \label{res}

\subsection{Radial pair correlation}

\begin{figure}
\centering
\includegraphics[width=\textwidth]{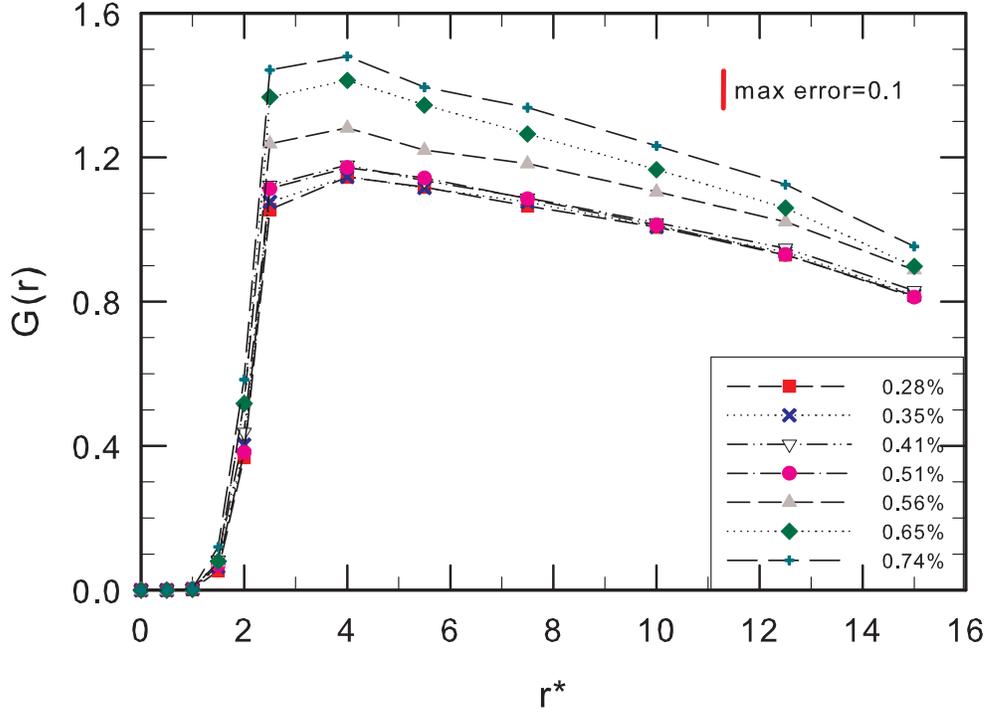}
\caption{Radial pair probability $G(r)$ as a function of dimensionless radius $r^*$. The different curves are for different bubble volume concentrations $\alpha$ (given in percent).} \label{fig:radial}
\end{figure}

The radial pair correlation  $G(r)$ was obtained for all bubble concentrations studied. Figure~\ref{fig:radial} shows $G(r)$ as a function of the normalized radius $r^*=r/a$, where $a$ is a mean bubble radius. As shown in figure~\ref{fig:bubdiam} the mean equivalent bubble diameter is within the range 4-5 mm. Therefore we can normalize with one mean bubble radius and we picked  $a=2$ mm for all concentrations. We observe in figure~\ref{fig:radial} that the highest probability to find a pair of bubbles lies in the range of few bubble radii  $r^*\approx$ 4 for all concentrations. The probability $G(r)$ of finding a pair of bubbles within this range increases slightly with increasing $\alpha$. For values $r^*<2$ one would expect that $G(r)=0$. However, in our experiments we found $G(r)\neq0$ for $r^*<2$, due to the fact that the bubbles are ellipsoidal and deform and wobble when rising.

We now estimate the error bar in the correlation function $G(r)$, originating from incomplete bubble detection, as seen from table \ref{t:imag_stat}. With increasing $\alpha$ the fraction of detected bubbles decreases. For $\alpha=0.28\%$ we detect $N_b\approx200$ so one would expect for $\alpha$=0.74\% to detect $N_b\approx(0.74/0.28)\times200\approx500$  but we are detecting only 110, $\approx$20\% of them. In order to investigate the reliability of the pair correlation function due to this loss of bubble detection, we studied a set of randomly distributed particles. We generated 500 particles at 6000 times and calculated the radial and pair correlation functions, obtaining $G(r)=1$ and $G(\theta)=1$ as expected for randomly distributed particles. Then we kept only 20\% of the particles and recalculated the correlation functions to see how much they deviate from 1. We found that the maximum deviation was less than 0.1 for both the radial  and the angular correlation functions. This is much smaller than the structure revealed in the $G(r)$ curve in figure~\ref{fig:radial}. We therefore consider the clustering with the preferred distance of $r^*=4$ as a robust feature of our data, in spite of incomplete bubble detection. In figure~\ref{fig:radial} this error bar corresponding to a maximum error of $G(r)$ at the most concentrated flow is also shown.

\subsection{Angular pair correlation}

\begin{figure}
  \centering
  \subfloat{\label{fig:angular_8}\includegraphics[width=0.55\textwidth]{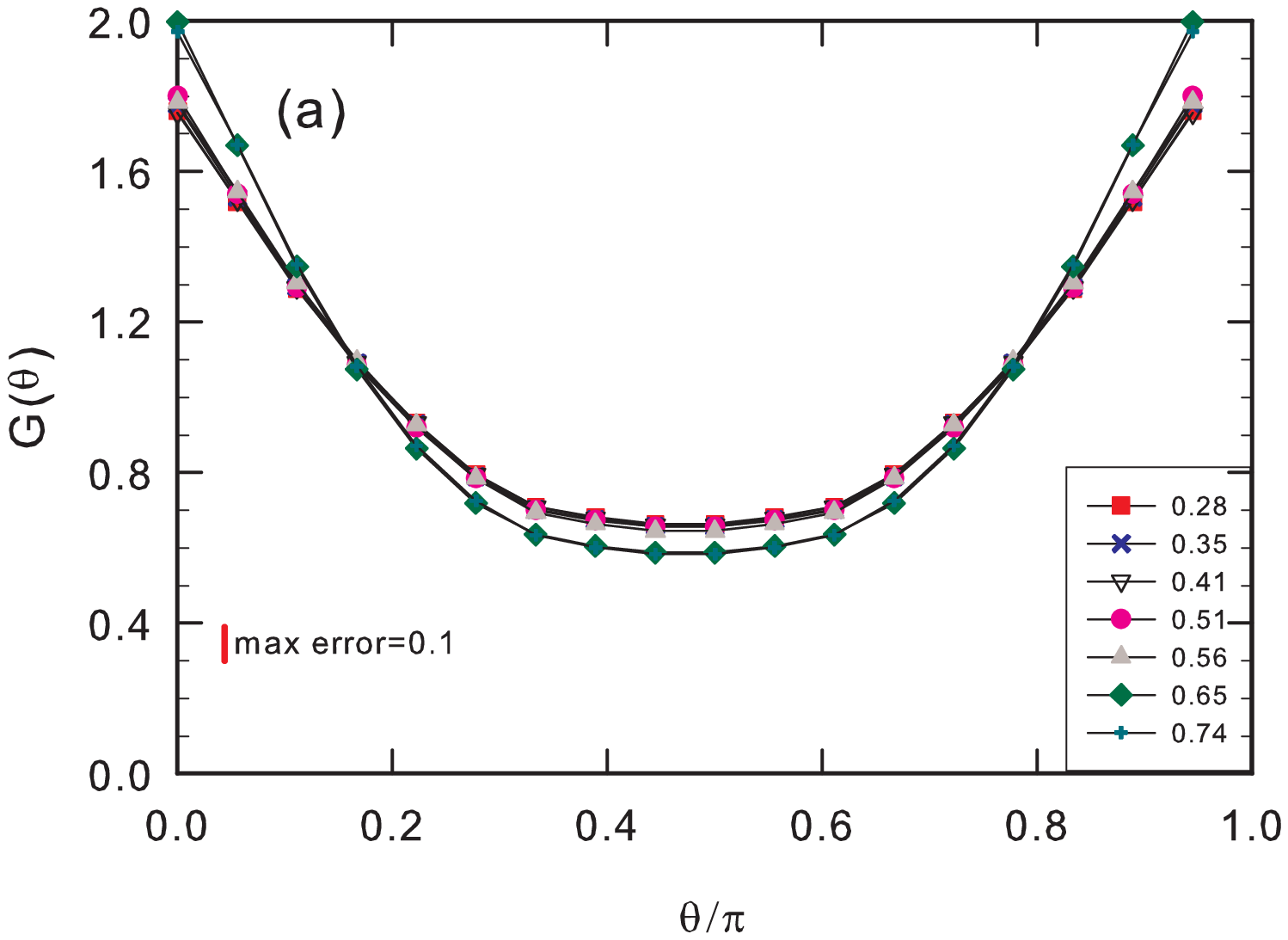}} \\
  \subfloat{\label{fig:angular_3}\includegraphics[width=0.55\textwidth]{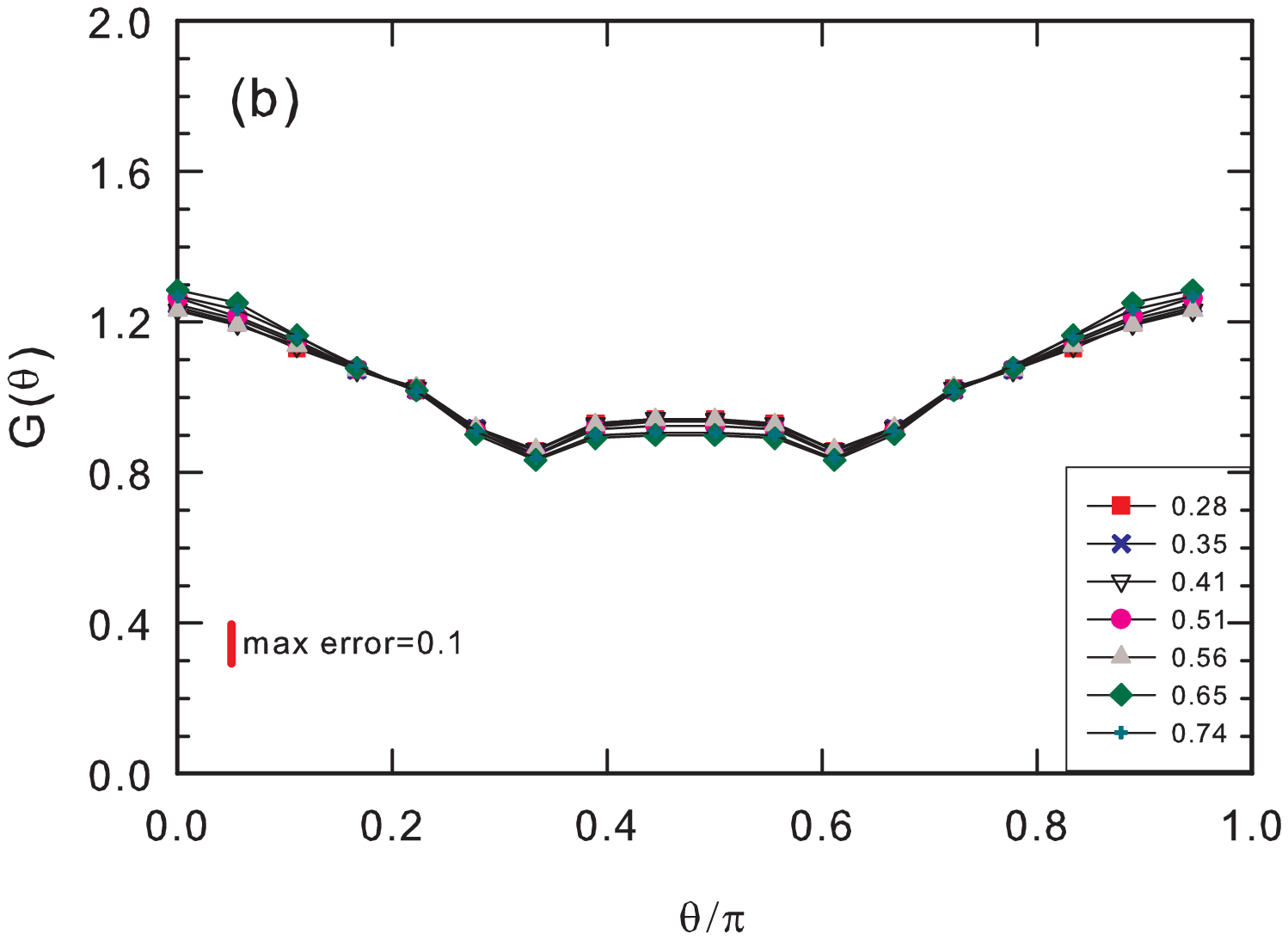}} \\
  \subfloat{\label{fig:angular_1}\includegraphics[width=0.55\textwidth]{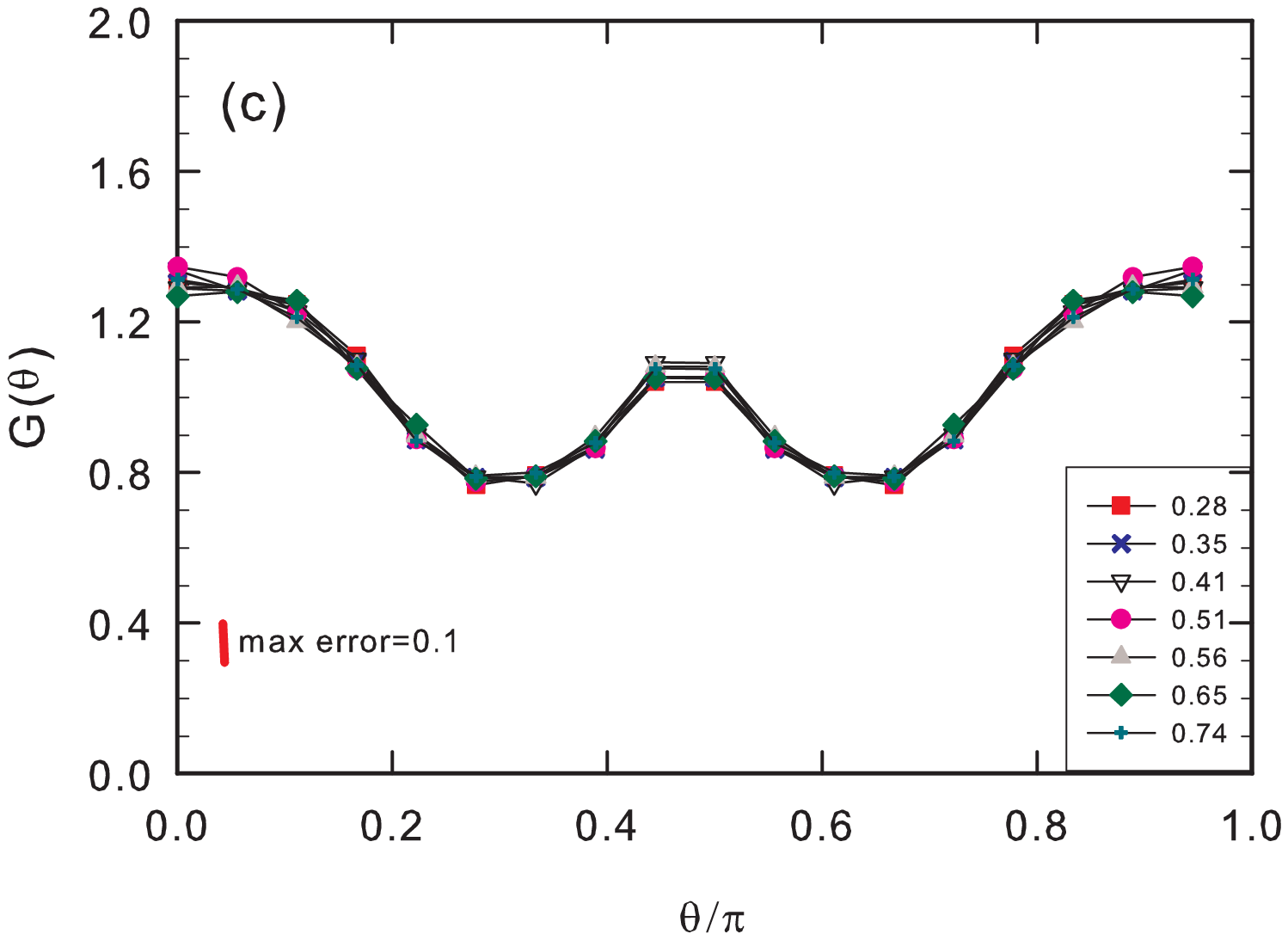}}
  \caption{Normalized angular pair probability $G(\theta)$ as a function of angular position $\theta/\pi$ for various bubble concentrations $\alpha$ (see insets) and three different bubble-pair distances: (a) $r^*=40$, (b) $r^*=15$, and (c) $r^*=5$. Maximum error bar for the angular correlation of 0.1 at $\alpha$=0.74\%.}
  \label{fig:angular}
\end{figure}

The orientation of the bubble clustering was studied by means of the angular pair correlation $G(\theta)$ using different radii for the spherical sector over which neighboring bubbles are counted. Figures~\ref{fig:angular_8}, \ref{fig:angular_3}, and \ref{fig:angular_1}  show the results for $r^*$=40, 15, and  5, respectively. The plots were normalized so that the area under the curve is unity. For all radii and concentrations,  pairs of bubbles cluster in the vertical direction, as one can see from the highest peaks at $\theta/\pi=0$ and $\theta/\pi=1$. The value of $\theta/\pi=0$ means that the reference bubble  (at which the spherical sector is centered) rises below the pairing one. For $\theta/\pi=1$ the reference bubble rises above the pairing bubble (see figure~\ref{fig:def_angle}). When decreasing the radius of the spherical sector,  i.e. when probing the short range interactions between the bubbles, we observe that a peak of the angular probability near $\pi/2$ starts to develop. The enhanced probability at this angle range is even more pronounced in figure~\ref{fig:angular_1} where the peak of $G(\theta)$ for horizontally aligned  bubbles is just slightly lower than that for vertical clustering. It is worthwhile to point out that the vertical alignment of the bubbles is very robust and is present from very large to small scales, as the angular correlation for different spherical sectors is always higher at values $\theta/\pi$=0 and 1 than at value $\theta/\pi$=0.5. The maximum error bar for the angular correlation for the most concentrated flow at $\alpha=0.74\%$, when 20\% of particles are detected, was 0.1 as explained above and is also shown in figure~\ref{fig:angular}. It is much smaller than the structure of the signal.

For comparison, we consider again the work of \cite{Trygg:II}, who found that pairs of bubbles have a higher probability to align vertically, though for  a much higher concentration ($\alpha=6\%$) than employed here. \cite{Trygg:II} found that the vertical alignment was not as robust as in our case, since with increasing $r^*$ the angular correlation at $0$ and $\pi$ became less dominant. Another significant difference between  the findings of our experimental work and their simulations is that  horizontal alignment was more pronounced with larger radii of the spherical sector and not when decreasing $r^*$. Our experimental results clearly show the main drawback and today's limitation when solving the flow at the particle's interface:  the simulations are still restricted to a small number of particles, which is not sufficient to reveal long range correlations.

\subsection{Interpretation of the clustering}
What is the physical explanation for a preferred vertical alignment of pairs of bubbles in pseudo-turbulence? Through potential flow theory, the mutual attraction of rising bubbles can be predicted \cite[][]{batchelorbook}, the application of potential theory to our experiments remains questionable \cite[][]{wijngaarden}, as we are in a statistically stationary situation where bubbles have already created vorticity.  Our findings are consistent with the idea that deformability effects and the inversion of the lift force acting on the bubbles are closely related to the clustering. \cite{mazzitelli:3}  showed numerically that it was mainly the lift force acting on point-like bubbles that makes them drift to the downflow side of a vortex in the bubble wake\footnote{See figure 2 in \cite{mazzitelli:3} sketching the dynamics.}. Furthermore, when accounting for surface phenomena,  \cite{trygg:IV} showed that the sign of the lift force inverts for the case of deformable bubbles  in shear flow so that a trailing bubble is pulled into the wake of a heading bubble rather than expelled from it. In such a manner vertical rafts can be formed. Experimentally some evidence of the lift force inversion has been observed by \cite{tomi} as lateral migration of bubbles under Poiseuille and Couette flow changed once the bubble size has become large enough. Numerical simulations of swarm of deformable bubbles without  any flow predicted a vertical alignment~\cite[][]{Trygg:II}. An alternative interpretation of the results, due to Shu Takagi (private communication (2009)) goes as follows: small, pointwise, spherical bubbles have a small wake, allowing for the application of potential flow. The bubbles then horizontally attract, leading to horizontal clustering. In contrast, large bubbles with their pronounced wake entrain neighboring bubbles in their wake due to the smaller pressure present in those flow regions, leading to vertical clustering. Further efforts are needed to identify and confirm the main mechanism---i.e., either lift or pressure reduction in the bubble wake--- leading  to  a preferential vertical alignment, for example through experiments with small, spherical, non-deformable bubbles as achieved by \cite{matsumoto} through surfactants  or with buoyant spherical particles.

\subsection{Average bubble rise velocity}

\begin{figure}
\centering
\includegraphics[width=0.9\textwidth]{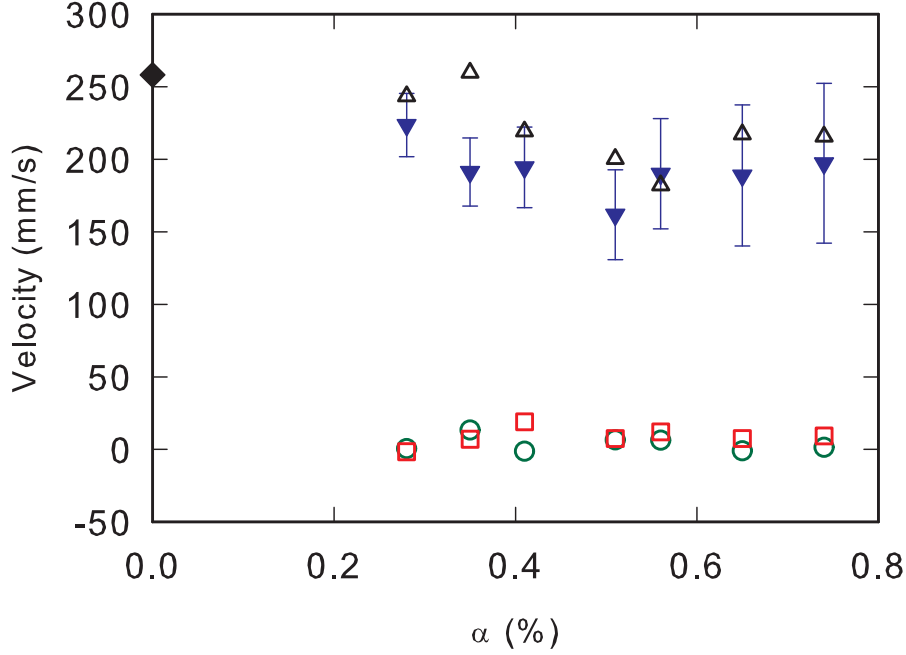}
\caption{(a) Mean bubble velocity as a function of bubble concentration $\alpha$. All three components are shown: $\circ$ $V_x$; $\square$ $V_y$; $\blacktriangledown$  mean bubble rise velocity $\overline{V}_z$; $\triangle$  mean bubble rise velocity obtained by particle tracking from single camera 2D images; $\blacklozenge$ terminal rise velocity for a single bubble with $d_b=3.9$ mm. The error bars were obtained by estimating the 95\% confidence interval for the mean.} \label{fig:mean_and_rms}
\end{figure}

Bubble velocities were calculated by tracking the bubble positions which were linked in at least three consecutive images. The mean  bubble rise velocity can thus be obtained as a function of the bubble concentration. Figure~\ref{fig:mean_and_rms} shows the  three components of the bubble velocity; the coordinate system corresponds to the one depicted in figure~\ref{fig:diagram_tunnel}. The terminal rise velocity of a single bubble with $d_b=3.9$ mm and with the same water-impurity condition is also shown in figure~\ref{fig:mean_and_rms}, it has a value of 0.26 m/s. A decrease in the mean bubble rise velocity with concentration is observed in our experiments within the experimental error showed in figure~\ref{fig:mean_and_rms}. The mean bubble rise velocity is 0.22 m/s for the most dilute case $\alpha=0.28\%$ and decreases until a value of 0.16 m/s for $\alpha=0.51\%$.

The interpretation of this finding is that in this parameter regime the velocity-reducing effect of the bubble-induced counterflow \cite[see][]{batchelor} and the scattering effect overwhelm the velocity-enhancing  blob-effect{\footnote{Originally suggested for sedimenting particles}} \cite[][]{brenner}, implying that a blob of rising bubbles rises faster than a single one.
For values of $\alpha\geq0.56\%$ the mean values are again larger, around 0.18 m/s. This increment of the mean rise velocity could be a result of our experimental error. As mentioned in section~\ref{sec:ptv}, the number of detected bubbles at higher concentrations decreases by a factor 3 as compared to the most dilute cases. To check whether this increment was coming from detection of blobs of bubbles rather than single ones, we did experiments with a single camera positioned perpendicularly to the flow. The 2D images were used to track bubbles manually making sure that the trajectories indeed corresponded to single bubbles. In figure~\ref{fig:mean_and_rms} the mean bubble rise velocity from the one-camera 2D analysis are plotted. A similar behavior is observed, first a decrease with concentration, followed by  a slight increase for the most concentrated flows, confirming the 3D PTV results.

\subsection{Bubble velocity distributions} \label{bub_vel_dist}
\begin{figure}
\centering
\includegraphics[width=\textwidth]{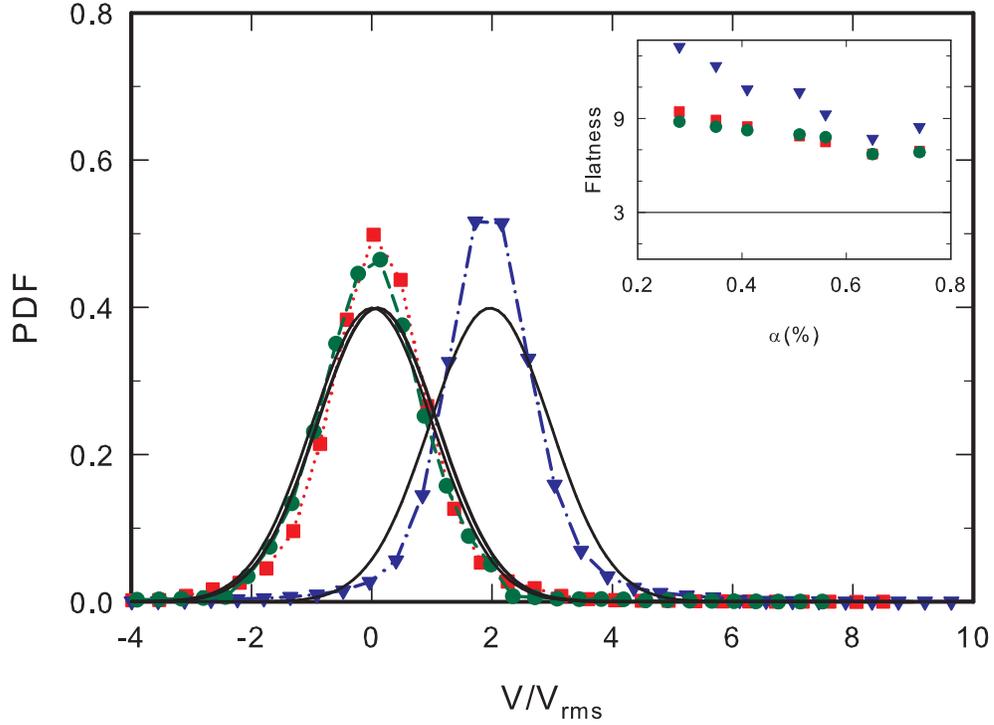}
\caption{Probability Density Functions for the three bubble velocity components for $\alpha=0.28\%$ normalized with the rms values of each component. $V_x$: dotted line with squares, $V_y$: dashed line with circles (both with $V_{mean}$ velocities as expected), $V_z$: dotted-dashed line with triangles. The respective black solid lines (without symbols) show a Gaussian with the same mean and width as the measured distributions. The inset presents the flatness of the distribution as a function of the concentration $\alpha$. The horizontal solid line in the inset represents the flatness for a normal distribution. Squares, triangles, and circles have the same meaning as in the main plot.} \label{fig:pdf_vel_3c}
\end{figure}

\begin{figure}
  \centering

  \subfloat{\label{fig:semilog_vx}\includegraphics[width=0.6\textwidth]{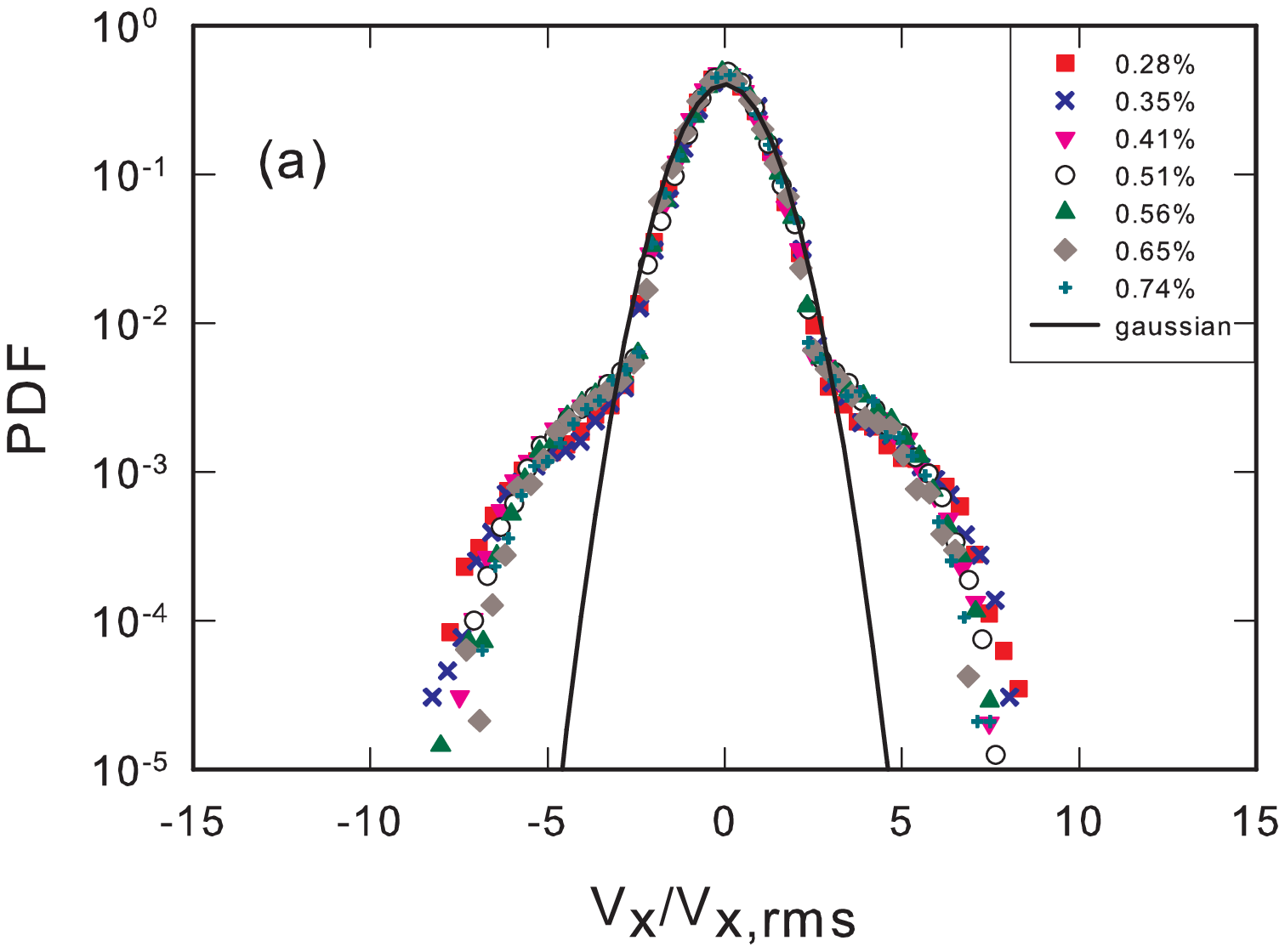}} \\
  \subfloat{\label{fig:semilog_vy}\includegraphics[width=0.6\textwidth]{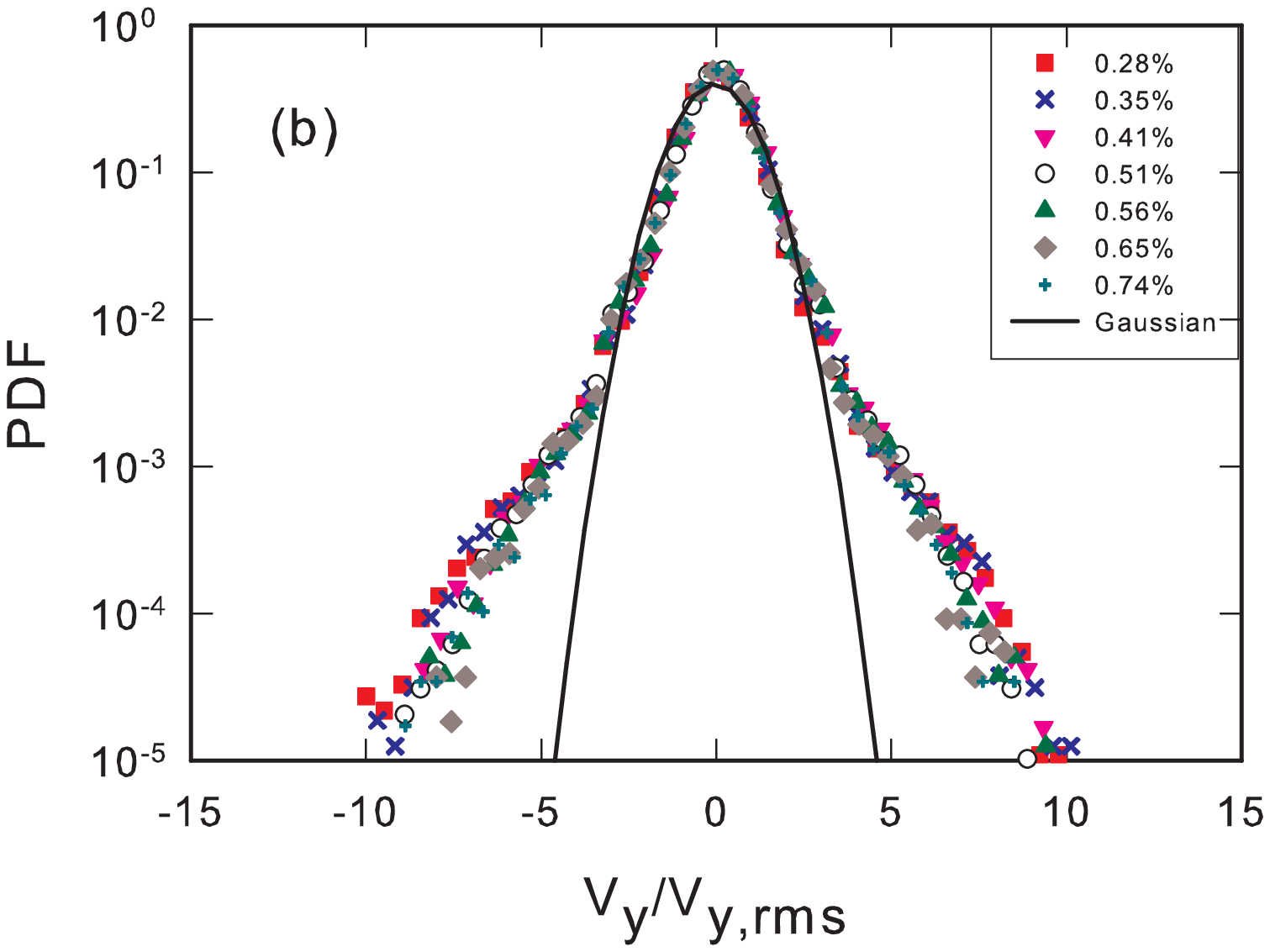}} \\
  \subfloat{\label{fig:semilog_vz}\includegraphics[width=0.6\textwidth]{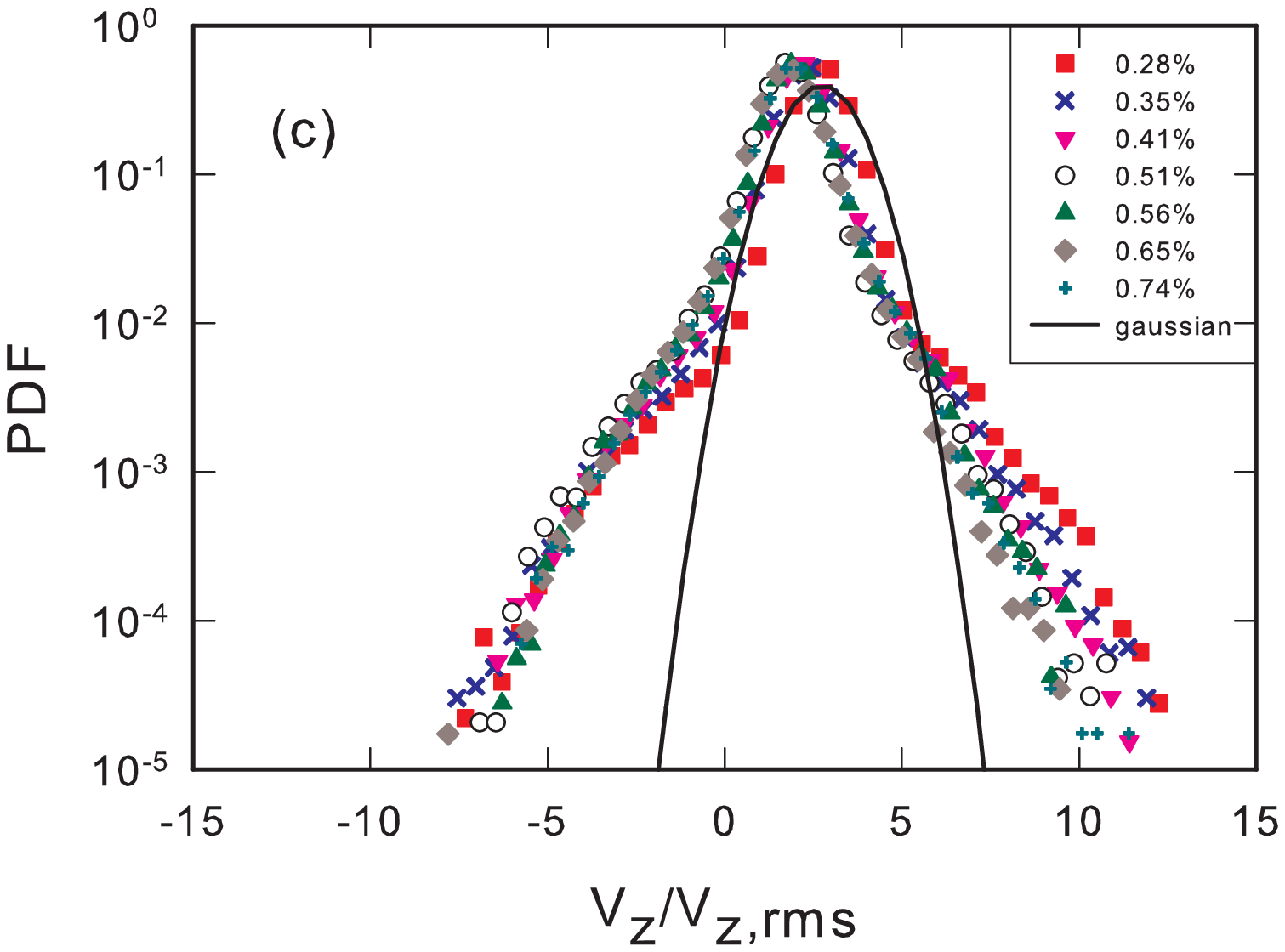}}

  \caption{Same Probability  Density Functions of the bubble velocity normalized with the rms values of each component as in figure~\ref{fig:pdf_vel_3c}, but now on a semi logarithmic scale to better reveal intermittency effects for various concentrations. As a reference, a Gaussian curve with the same mean and standard deviation as for  the distribution with $\alpha=0.28 \%$ is plotted as a solid line.}
  \label{fig:semilog}
\end{figure}

Figure~\ref{fig:pdf_vel_3c} shows the PDF of all velocity components at the most dilute concentration ($\alpha=0.28\%$). The number of data-points used for calculating the PDFs was larger than 9$\times$10$^4$ for the highest concentration $\alpha$=0.7\% and of order 3$\times10^5$ for the most dilute case. Even for the most concentrated flow, the number of data-points was large enough to assure the statistically convergence of the PDFs.  All PDFs show non-Gaussian features, as nicely revealed in the semi-log plot of the PDFs (figure~\ref{fig:semilog}).

In order to quantify the non-Gaussianity of the PDFs, the flatnesses of the distributions were calculated. Their values are shown in the inset of figure~\ref{fig:pdf_vel_3c} and are within the range 6$-$13. The flatness of the vertical component has always the highest values in  the whole range of void fraction.

The effect of the concentration on the PDFs is also shown in figure~\ref{fig:semilog}.  As in that figure, the PDFs are shown on a semi-logarithmic scale so that the deviation from the Gaussian-like shape become more visible, it is revealed that there is no substantial change in the shape of the distributions with increasing bubble concentration.

We would like to compare the non-Gaussianity of the PDFs (``intermittency") found here with a comparable system, namely Rayleigh-B\'enard convection as the analogy between bouyancy driven bubbly flows and free thermal convection is interesting \cite[see][]{magnaudet:thermal}. In Rayleigh-B\'enard convection a fluid between two parallel plates is heated from below and cooled from above, see \cite{lohse_ahlers} for a recent review. Prominent  coherent structures in this system are  thermal plumes, which are fluid particles either hotter or colder than the background fluid. Due to the density difference with the background fluid, hotter plumes ascend and colder ones descend. The system is solely buoyancy driven. Particularly, for large $Pr$ the plumes keep their integrity thanks to the small thermal diffusivity, so that the similarity with pseudo-turbulence is appealing\footnote{We stress however that of course there are differences between the two systems which have been pointed out by \cite{magnaudet:thermal}}.

Does the statistics of the velocity fluctuations in Rayleigh-B\'enard share a similar behavior with that of bubbles in pseudo-turbulence?
In Rayleigh-B\'enard convection, the PDF of the vertical velocity fluctuations of the background fluid---i.e. the central region of the cell---exhibits a Gaussian distribution \cite[][]{daya}. \cite{qiu}, and \cite{sun} measured the vertical velocity fluctuations in the region where the plumes abound, they found that the PDF still follows a Gaussian function. Those measurements, carried out for $Pr\approx4$, indicate clearly that the PDF of the plume velocity fluctuations can be described by a Gaussian function, which differs significantly from the statistics of the bubble velocity in pseudo-turbulence. A possible reason of this difference could be that buoyancy in pseudo-turbulence is much stronger than that of the plumes in Rayleigh-B\'enard system.

\subsection{Energy spectra of pseudo-turbulence}\label{sec:spectrum}

\begin{figure}
\centering
\includegraphics[width=0.8\textwidth]{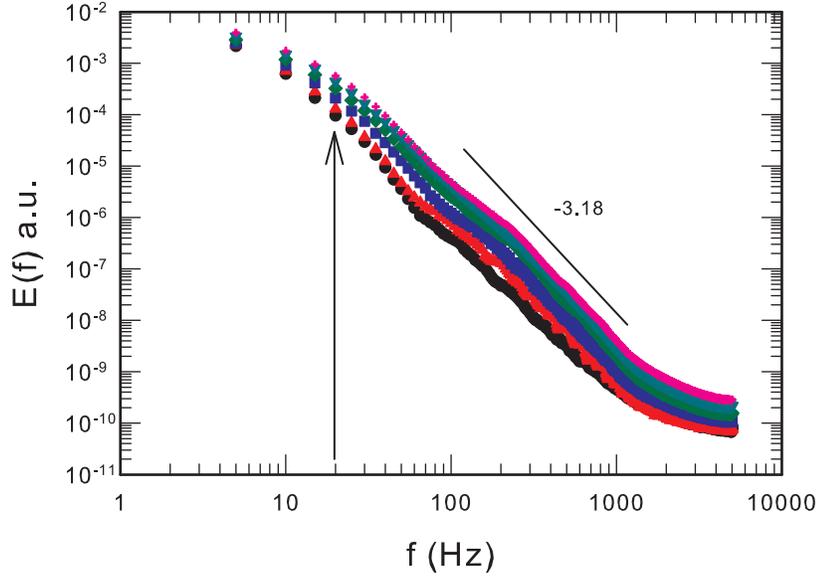}
\caption{Liquid energy spectra for different void fractions. $\bullet$ $\alpha$=0.2\%, $\blacktriangle$ $\alpha$=0.4\%, $\blacksquare$ $\alpha$=0.8\%, $\blacklozenge$ $\alpha$=1.3\%, $\blacktriangledown$ $\alpha$=1.7\%, $+$ $\alpha$=2.2\%. The arrow shows the onset frequency of the scaling.} \label{fig:spectra}
\end{figure}

Figure~\ref{fig:spectra} shows the energy spectra for all gas fractions. As it can be seen, the slope of the energy spectrum hardly depends on the volume fraction---all curves show a slope of about $-$3.2.  We stress that this scaling behavior is maintained for nearly 2 decades, much wider than it had been reported in prior observations \cite[][]{lance,kazu,riboux} of this steep slope of pseudo-turbulence spectra. As it was mentioned in section~\ref{sec:cta}, the way  the power spectrum was calculated in this investigation differs from previous ones in two aspects: First, the indicator function has been measured by means of the optical fibers and second, an energy spectrum has been calculated for all individual liquid segment, before the final spectrum is obtained through averaging.

\begin{figure}
\centering
\includegraphics[width=0.65\textwidth]{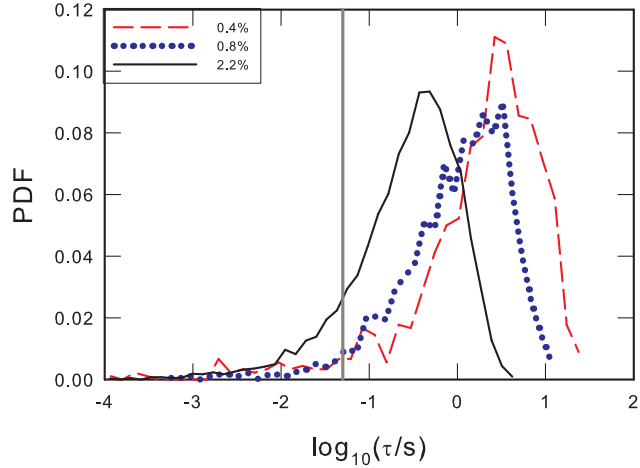}
\caption{Distribution of the time duration of the parts containing liquid information in the CTA-signal used to calculate the spectrum for three different bubble concentrations. The solid vertical line corresponds to the onset frequency of the scaling in figure \ref{fig:spectra}, see arrow in that plot.} \label{fig:hist_spec}
\end{figure}

One wonders whether the duration of our interrupted time series is large enough to resolve the low frequency part of the spectrum: if the duration of these segments is too short,  then indeed the low frequencies in the power spectrum can not be  resolved.  On the other hand,  if the duration of the liquid segments is large enough, then all frequencies in the spectrum are well resolved. Figure~\ref{fig:hist_spec} shows the distribution of the logarithm of the non-interrupted time series duration for three different concentrations. For $\alpha$=2.2\% (the most concentrated bubbly flow with more bubble-probe interaction, thus shorter liquid segments) around  90\% of the segments used to construct the spectrum have a duration larger than 0.05 s. Comparing with figure~\ref{fig:spectra}, we can see that for frequencies higher than 20 Hz we have resolved the inertial range, where the slope is $\approx-3.2$. Thus the measurement of the spectra is consistent.

Why does the slope differ from the Kolmogorov value $-5/3$? One might expect a different scaling in pseudo-turbulence, as the velocity fluctuations are caused by the rising bubbles and not by the Kolmogorov-Richardson energy cascade, initiated by some large scale motion. The difference between these two scalings is not yet completely understood, but there are some hypothesis on its origin. One possible explanation for the different scaling in pseudo-turbulence was given by \cite{lance}. They argued that eddies from the bubble wake are immediately dissipated before decaying towards smaller eddies, which would lead to the $-5/3$ scaling. They derived a $-3$ scaling,  balancing the  spectral energy, and assuming that the characteristic time of spectral energy transfer is larger than those of dissipation and production. More evidence that wake phenomena are related to the $-3$ scaling has been given by \cite{risso1} and by  \cite{roig}. They showed that bubbles' wakes decay faster in pseudo-turbulence than in the standard turbulent case with the same energy and integral length scale. They also proposed a spatial and temporal decomposition of the fluctuations in order to gain more insight into the different mechanisms. In their experiments they used a fixed array of spheres. Very recently, \cite{riboux} measured the spatial energy spectrum in pseudo-turbulence by means of PIV. They measured the spectrum just immediately after a bubble swarm has passed and obtained a $-$3 decay for wavelengths larger than the bubble diameter (2 mm $<l_c<$7 mm). For wavelengths smaller than the bubble diameter they found that the spectrum recovered the $-5/3$ scaling. Their findings showed that the scaling is independent on the bubble diameter and void fraction in their range of parameters investigated ($0.2\%\leq \alpha \leq 12\%$ and $d_b= 1.6-2.5$ mm), which we also find. Their conclusion is that the $-3$ scaling is only a result of the hydrodynamic interactions between the flow disturbances induced by the bubbles. Their arguments for this statement are that the scaling appears for wavelengths larger than the bubble diameter and that a different scaling was found for smaller wavelengths. It is worthwhile to emphasize that in our case, we measured \emph{within} the swarm where production is still maintained and steady. The $-$3 scaling in our measurements is in the range of 8 mm  down  to hundreds of micrometers\footnote{estimated by considering the  mean bubble rise velocity and the starting and ending frequencies of the scaling of the spectrum in figure~\ref{fig:spectra}}, thus for lengths not only larger than $d_b$ but also for those up to one order of magnitude smaller than it. This supports that the dissipation of the bubble wake is involved and still a valid explanation for the $-$3 scaling  as proposed by \cite{lance}. In any case, the papers by \cite{kazu,roig,risso1,riboux} and the present one show that the $-$3 scaling is typical for pseudo-turbulence.

One further result supporting this idea is the one obtained by \cite{mazzitelli:2} who performed numerical simulations of microbubbles in pseudo-turbulence modeling them as point particles. Their DNS treated up to 288000 bubbles, where near-field interactions were neglected, thus wake mechanisms can not be accounted for,  and effective-force models were used for the drag and lift forces. They obtained a slope of the power spectrum close to $-5/3$ typical for the turbulent case. This gives evidence that the bubble's wake---missing in the point particle approach--- and its dissipation play a very important role for the $-$3 scaling of the energy spectrum in pseudo-turbulence.

\section{Summary} \label{conclusions}

We performed experiments on bubble clustering using 3D-PTV. This is the first time that the technique is used to investigate  bubbly flows in pseudo-turbulence at very dilute regimes ($\alpha<1\%$). Bubble positions were determined  to study bubble clustering and alignment. For that purpose the pair correlation function $G(r,\theta)$ was calculated. As the radial correlation $G(r)$ shows, pairs of bubbles  cluster within few bubble radii $2.5<r^*<4$. Varying the bubble concentration does not have any effect on the clustering distance. The angular pair correlation $G(\theta)$ shows that a robust vertical alignment is present at both small and large scales, as it is observed when varying the radius of the spherical sector ($r^*$=40, 15, and  5). Decreasing the radius of the spherical sector shows that horizontal clustering also occurs, as the peak of the angular correlation around $\pi/2$ starts to grow with decreasing values of $r^*$.

Probability density functions of bubble velocity  show that all components of bubble velocity behave differently from Gaussian. The implementation of this non-intrusive imaging technique assures enough data points to obtain  convergence in PDFs. The improvement achieved in the number of data-points, compared with previous experimental investigations,  is of order 10$^2$. Furthermore, this allowed us to show the intermittent signature that bubble distributions have for all components. The flatness values for these velocity distributions are in the range of 6$-$13.  The distribution of the rise velocity showed the highest values of  flatness, e.g. $\approx13$ at $\alpha=0.28$\%. The non-Gaussianity can be a result of the cluster formation mechanism, where the rise velocity of single bubbles are affected by the faster collective motion of clusters. However further investigations are needed to fully understand its origin.

We have shown that the power spectrum in  pseudo-turbulence ($b=\infty$) decays exponentially with a slope near $-3$ which is consistent with the theoretical scaling that \cite{lance} derived and supporting the hypothesis that bubble wake mechanisms are closely related to it. We have shown that the implementation of Phase Sensitive CTA for studying bubbly flows is of great advantage, allowing for a direct recognition and discard of bubble-probe interactions.

The next step of our research will be to investigate bubble clustering for $b<<1$, where turbulent effects become dominant. Another line of research is to analyze pseudo-turbulence with smaller bubbles (e.g. achieved by surfactants \cite[][]{matsumoto}), to study the effect of the length scale of the bubble on the spectra.

\section*{Acknowledgements}
We thank specially Gert-Wim Bruggert, Martin Bos, and Bas Benschop for their invaluable help in the experimental apparatus. We thank: Lorenzo del Castello, Beat L\"uthi, and Haitao Xu for all the help and discussions regarding the PTV system, Fr\'ed\'eric Risso, Veronique Roig, and Roberto Zenit for the discussion on energy spectra in pseudo-turbulence, Rob Mudde for help and discussion on probe related issues, Shu Takagi, Yoichiro Matsumoto, Bert Vreman. Special thanks for one of the referees for his/her constructive and helpful comments on the pair correlation function. This research is part of the Industrial Partnership Programme: \emph{Fundamentals of heterogeneous bubbly flows} which is funded by the Stichting voor Fundamenteel Onderzoek der Materie (FOM).

\bibliographystyle{jfm}
\bibliography{Martinez09}

\end{document}